\documentclass[12pt,preprint]{aastex}

\newcommand{\ha}{H$\alpha$}

\newcommand{\oiii}{[O~{\sc iii}]~5007~\AA}

\newcommand{\degree}{$^{\circ}$~}
\def\vhel{\ifmmode{V_{{\rm HEL}}}\else{$V_{{\rm HEL}}$}\fi}
\def\vsys{\ifmmode{V_{\rm sys}}\else{$V_{\rm sys}$}\fi}
\def\kms{\ifmmode{~{\rm km\,s}^{-1}}\else{~km~s$^{-1}$}\fi}
\def\thetc{$\theta^{1}$Ori C}
\newcounter{no1}

\shortauthors{Graham et al.}
\shorttitle{Interaction Zone in Binary Proplyd LV 1}

\begin{document}

\title{MERLIN radio detection of an interaction zone within a
binary Orion proplyd system.}
\author{M. F. Graham, J. Meaburn, S. T. Garrington and T. J. O'Brien}
\affil{Jodrell Bank Observatory, University of Manchester,
Macclesfield, Cheshire, SK11 9DL, UK}
\email{mgraham@ast.man.ac.uk}
\author{W. J. Henney}
\affil{Instituto de Astronom\'\i a, Universidad Nacional
  Aut\'onoma de M\'exico, Campus Morelia, Apartado Postal 3--72
  (Xangari), 58089~Morelia, Michoac\'an, M\'exico}
\author{C. R. O'Dell}
\affil{Department of Physics and Astronomy, Box 1807-B, Vanderbilt
University, Nashville, TN 37235}

\label{firstpage}

\begin{abstract}
  Presented here are high angular resolution MERLIN 5 GHz (6 cm)
continuum observations of  the binary proplyd system, LV 1 in the
Orion nebula, which consists of proplyd 168--326SE and its binary
proplyd companion  168--326NW (separation $0.4\arcsec$). Accurate 
astrometric alignment allows a detailed comparison between these data
and published \emph{HST} PC H$\alpha$ and \oiii\ images.
 
Thermal radio sources
 coincide with the two proplyds and originate in the ionized
  photoevaporating flows seen in the optical emission lines.
Flow velocities of $\approx\ 50 \kms$ from the ionized proplyd
surfaces and $\geq 100 \kms$ from a possible micro-jet 
have been detected using  the Manchester Echelle
spectrometer.

  A third radio source is found to coincide with a region of extended,
high excitation,  optical line emission that lies between the binary 
proplyds  168--326SE/326NW . This is modelled as a 
bowshock due to the collision of the
 photoevaporating flows from the two proplyds. Both a thermal and a
non-thermal origin for the radio emission in this collision zone are
considered.
 
\end{abstract}

\section{Introduction}

It is now well established that the compact, highly ionized
gaseous knots (LV 1-6) near the Trapezium stars in M42, discovered by
Laques \& Vidal (1979) contain young low mass stars (YSOs)
still partially cocooned in their primaeval material. See Figure 1 for their
locations.  Many similar objects were
found in the same vicinity at radio wavelengths (Churchwell et
al. 1987; Garay 1987; Garay, Moran, \& Reid 1987; Felli et al. 1993a;
Felli et al. 1993b). 
The LV knots are particularly interesting  both 
because of their close proximity
to the Sun,
permitting detailed observation, and because of 
their extreme local environment. 
This environment is swept by the energetic particle wind
of the nearby O6 star, \thetc\ and the LV knots are irradiated by its 
Lyman (and near UV) photons. 
Their outside surfaces are
consequently highly ionized, though shielded form the direct wind from
\thetc\ by a standoff bowshock (Figure 3) in the photoevaporated flow.

\emph{HST} imagery revealed the structure of these systems, now
known as `proplyds' (O'Dell, Wen, \& Hu 1993),
with startling clarity (O'Dell \& Wong 1996; Bally et al. 1998). The
central stars, found originally on a 2 $\mu$m IRCAM `engineering'
image with UKIRT by Meaburn (1988) and subsequently by McCaughrean \&
Stauffer (1994) are seen to be embedded in dense  cocoons/disks of
dusty  molecular gas. Photoevaporated flows from the ionized surfaces
form cometary tails pointing away from \thetc\ as they meet the UV
flux of this star. Supersonic
flow velocities of $\geq$ 50\kms\ were found (Meaburn 1988; Meaburn
et al. 1993; Massey \& Meaburn 1995) using the Manchester Echelle
spectrometer (MES - Meaburn et al. 1984) 
on a variety of telescopes. The origin of these ionized flows has been
modelled, initially as a two-wind interaction (Henney et al. 1996,
1997) and more recently as the interaction of the flow from the
proplyd disk with the ionizing Lyman photon field of \thetc\
and the nebula (Johnstone, Hollenbach, \& Bally 1998; Henney \& Arthur
1998; Henney \& O'Dell 1999; St{\"o}rzer \& Hollenbach 1999). The
two-wind interaction is now thought to manifest itself much further
from the proplyd disk in the form of the standoff bowshocks (one of
which is shown in Figures 1 and 3) observed by Hayward, Houck and
Miles (1994),   McCullough et al. (1995) and  Bally et al. (1998).
The flow from the proplyd itself is
most likely caused  by photoevaporation of the disk by either ionizing
(EUV) or non-ionizing (FUV) photons from \thetc, depending on the
proplyd's proximity to it (Johnstone, Hollenbach, \& Bally 1998).
The process of photoevaporation, photoionization, and acceleration of the 
disk gas down the surrounding density gradient is found to produce
flow velocities of only 2-3 times the sound speed of $\sim$10\kms\
(Dyson 1968), which does not fully account for the velocities observed
in the longslit spectra. These longslit spectra were complemented by
the Fabry-Perot work of O'Dell et al. (1997), which compared velocity
data with \emph{HST} images and suggested the presence of a micro-jet
associated with one of the proplyds.

The presence of predominantly monopolar jets from the YSOs in many of
these proplyds was first suggested
by the $\geq$ 100\kms, spatially compact ($\leq$ 1 arcsec across) 
spikes or knots on the 
MES position-velocity (pv) arrays of Meaburn et al. (1993) and Massey
\& Meaburn (1995).  This has now been confirmed through direct
\emph{HST} imagery by Bally, O'Dell, \& McCaughrean (2000), who find
more than 20 jets associated with various proplyds.  This apparent
ubiquity of jets and the fact that it is difficult to get high
velocity flows from density gradients suggest that most collimated
high velocity features are jets.  \emph{HST} imagery, as a consequence
of its high angular resolution, along with high resolution near
infra-red imagery (Petr 1998),  has also revealed the binary nature of
a small fraction of the proplyds.

The binary structure of the prominent proplyd originally designated
LV 1 when unresolved in ground-based observations is of particular
interest. It was first resolved into two sources in the 2 cm
VLA A-array radio maps of Felli et al. (1993a) (Sources 5
and 20 in their Tables 2 and 3).  The higher resolution \emph{HST}
image shows that this system is formed by two 
proplyds separated by 0.4\arcsec, which are catalogued as 168--326N
and as 168--326S, by O'Dell \& Wen (1994). In the subsequent full
catalogue of stars and compact objects of O'Dell \& Wong (1996), it
appears as a single proplyd, 168--326, corresponding to Felli et
al.'s (1993a) source 20.  However, the near infra-red 0.13\arcsec\
resolution binary star survey of Petr (1998) identifies 168--326 as a 
binary system consisting of two components that are referred to as
168--326E and 168--326W.  Bally et al. (1998), who used the same
\emph{HST} observations as used here, label the two components as
168--326 and 168--326a (their figure 2) and mention the presence of a
bright arc lying between the sources.  Here, the proplyds
are referred to as 168--326SE and 168--326NW, which is the clearest
form of designation.  In the present
paper the discovery is reported of the 
interaction zone between the flows from 168--326SE and 
326NW in both the radio (at 5 GHz with MERLIN) and optical (\ha\ and
\oiii\ with the \emph{HST}) domains. The collision of two
photoevaporated flows is explored to explain the nature of this
interaction.  Further kinematical and morphological evidence for a jet 
in LV 1 is also presented.

\section{Observations}

\subsection{Imagery}

\subsubsection{Radio}

MERLIN observations of the Orion nebula were made between
1998 December 21 and 1999 February 15 at 5 GHz
(6 cm) with a bandwidth of 15 MHz.  The observations, taken over 8 days,
alternated every 10 minutes between a phase calibrator, 0539-057 
and the target
itself.  3C286 was used to get a flux for a point source at the time 
of the observations.  The phase calibrator was used in order to
remove atmospheric density variations from the target source data and
also to register the astrometric alignment of the images to an accuracy
approximately 0.010\arcsec.
Initial data editing and amplitude calibration were done using the
MERLIN-specific d-programs.  Subsequent phase calibration was then
carried out in the National Radio Astronomy Observatories' (NRAO) AIPS 
package.  The resulting data were CLEANed using the AIPS task IMAGR
and finally restored with a circular beam of full width at
half-maximum, FWHM = 0.080\arcsec. More details of the reduction
of the radio data can be found in a parallel paper on LV 2
(Henney et al. 2001).

\subsubsection{Optical}

The optical observations presented here were obtained with
the \emph{HST}'s Planetary Camera (PC) under J. Bally's observing
program, GO 5469.  The images used were obtained at the target named
NGC1976-LV3 through the F656N (\ha) and F502N (\oiii) filters.  For
each filter there were three images at this position, each lasting 60 s,
for the \ha\ filter and 100 s for the \oiii\ filter.  In both cases, the
three images were combined using the NOAO IRAF package's CRREJ
routine, which removes cosmic rays during the combining process.  The
images were then rotated by an amount such that north was at the top
of the image and east was at the left, using the image header keyword
ORIENTAT, which gives the position angle of the y axis of the PC
detector projected on the sky, in degrees east of north.

Images in the F658N and F547M filters were also used to make an exact
 flux calibration 
using the method of O'Dell and Doi (1999), which corrects for
 underlying continuum
and contamination by adjacent lines.

\subsubsection{Alignment of Radio and Optical Images}

The process of phase referencing MERLIN images results in an absolute
astrometric calibration that is accurate to approximately 0.010\arcsec\ 
and tied to the radio reference frame (ICRF).  However the absolute
co-ordinate system that is built in to \emph{HST} images is known
to be inferior to this by an order of magnitude.  Therefore, alignment 
of the optical and radio images was done by shifting the co-ordinate
system of the optical images and taking the radio co-ordinate system
to be correct.  Ideally, the optical images could be re-calibrated
empirically using stars in the field with \emph{HIPPARCOS} positions,
of which there are three.  Unfortunately, these three stars,
$\theta^1$Ori A (HIP 26220), \thetc\ (HIP 26221), and
$\theta^1$Ori D (HIP 26224) are badly saturated in the \emph{HST} PC
images, meaning that their positions on the PC CCD chip cannot be
reliably determined.

In the vicinity of \thetc, MERLIN clearly detected all six LV
knots at 5 GHz. These six proplyds are also obvious on the
\emph{HST} images (figure 2 of Bally et al. 1998). A comparison of the
centroided optical and radio positions as given by their respective
co-ordinate systems leads to a consistent difference between the
two. As a result, this difference of a few hundred milliarcseconds was
applied to the co-ordinate system of the \emph{HST} images, in order
to correct and align them with the MERLIN image.  It is not
unreasonable to expect that there is a real difference in the position
of the optical and radio emission from these objects, which puts this
method into question.  However, this 
difference is likely to be a lot less than the gross error of a few
hundred milliarcseconds in the \emph{HST}'s astrometric calibration.
Also, this real separation in optical and radio position would be
expected to be along the line joining the proplyd to the ionizing
source (\thetc). Therefore, the fact that the five proplyds
used for the alignment  surround \thetc\ and are oriented at
various position angles means that the real difference is somewhat
preserved. In effect,  the difference in radio and optical positions
has been minimised as opposed to removed. It should be noted that LV 6
was not used because its radio position was hard to determine, due to
its apparently complicated structure.  $\theta^{1}$Ori A is a strong
radio source, but could not be used in the alignment as its radio
position was found to be significantly displaced from its optical
position.  The final astrometric calibration was found to be
consistent with that of Henney et al. (2001).

\subsection{Echelle Spectroscopy}

The Trapezium region of the Orion nebula, shown in Figure 1, has been
explored spectroscopically with longslit spectra obtained with the MES
(Meaburn  1988; Massey \& Meaburn 1993; Meaburn et al. 1993; Massey \&
Meaburn 1995).
%
\begin{figure}
\resizebox{\textwidth}{!}
   {\includegraphics{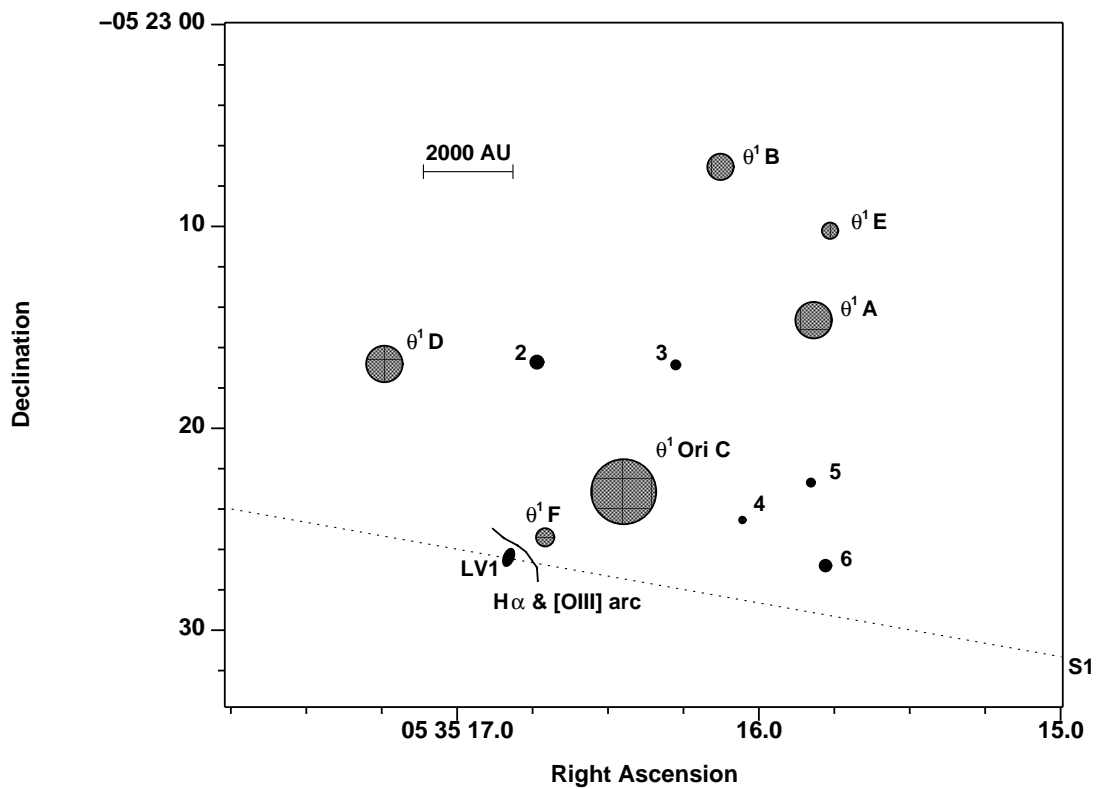}}
\caption{Identification chart for the Trapezium region.  The
$\theta^1$Ori stars are shown as grey circles and labelled with their
$\theta^1$Ori designation. Proplyds are shown as black circles and are
labelled with their LV number.  LV 1 is shown as an ellipse to indicate 
its extended binary nature.  The position of the arc/bow shock that
shields the LV 1 binary system from the fast wind of \thetc\ is
also shown, along with the position of the MES slit, S1.}
\end{figure}
%
The pv array of \oiii\ profiles
shown in Figure 2 was taken at the slit
position referred to as S1 in Massey \& Meaburn (1995) with the MES on
the 2.5m Isaac Newton Telescope.
%
\begin{figure}
\resizebox{\textwidth}{!}
   {\includegraphics{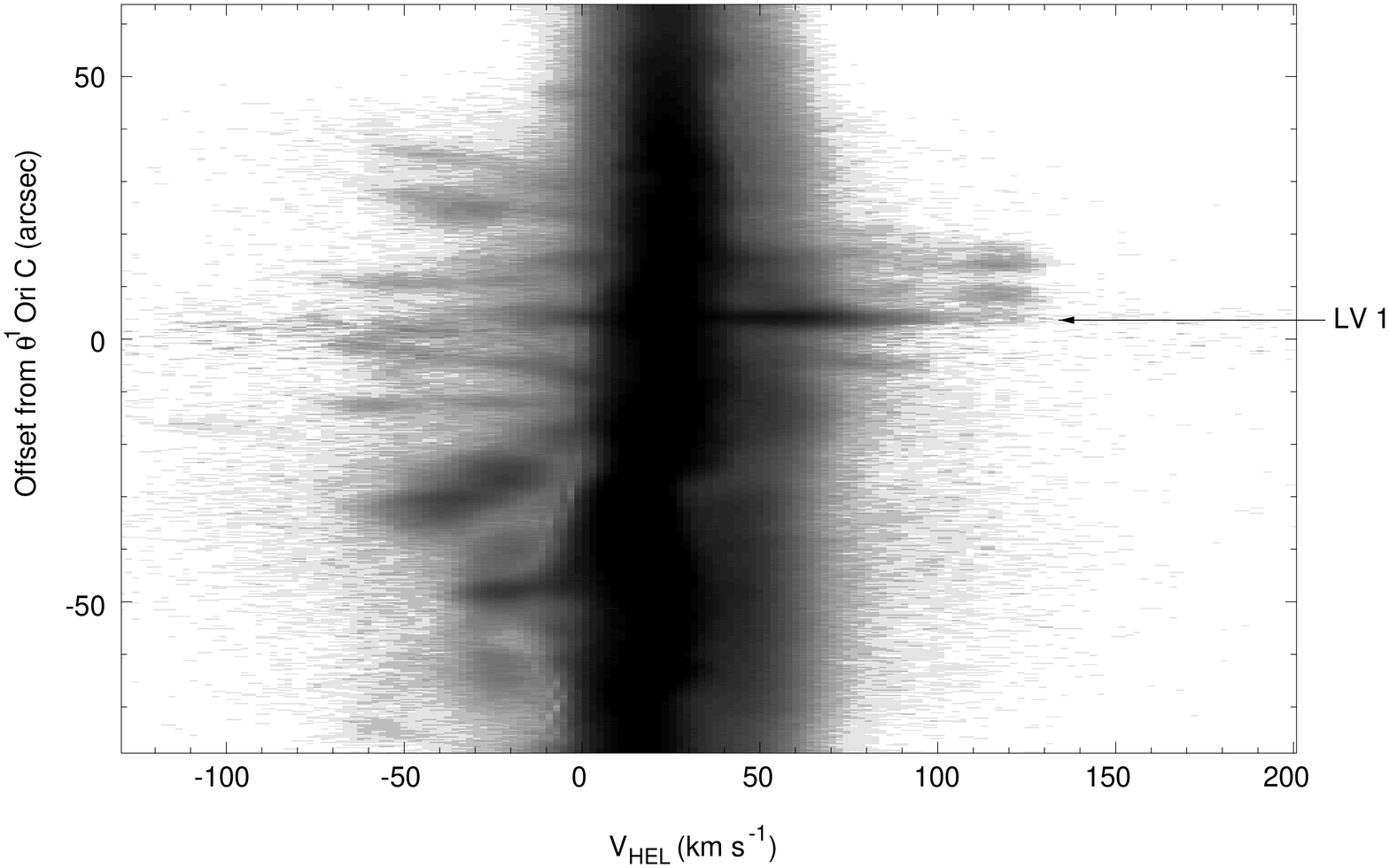}}
\caption{\oiii\ pv array of LV 1 taken with the Manchester
echelle spectrometer at slit position S1.  The position of LV 1 on this 
slit length is indicated.  A positive offset is eastwards along the
dotted line shown in Figure 1.}
\end{figure}
%
The slit position is shown in Figure 1 and the full observational setup
for the data can be found in Massey \& Meaburn (1995).  Stellar
continuum from \thetc\ has  now been subtracted from this spectrum,
leaving some residual noise, but also allowing the emission from LV 1
to be more easily appreciated. 

The emission from LV 1 is identified in Figure 2.  It is most intense
out to velocities of $\pm50\kms$ with respect to the nebula and
extends more faintly to $\approx$ 100\kms\ with respect to the
systemic 
heliocentric radial velocity, \vsys\ = 25\kms. 
Such a high speed `spike' in the pv array is
indicative of a jet and although one was not originally identified
in the \emph{HST} imagery of Bally, O'Dell, \& McCaughrean (2000),
Garc{\'i}a-Arredondo, Henney \& Arthur (2001) present evidence of 
a micro-jet from 168-326SE at a position angle of +85$^{\circ}$.  This
jet is shown in Figure 3 to be predominantly monopolar, although there 
is a hint of an approaching component in the PV array in Figure 2. 
%
\begin{figure}
\resizebox{\textwidth}{!}
   {\includegraphics{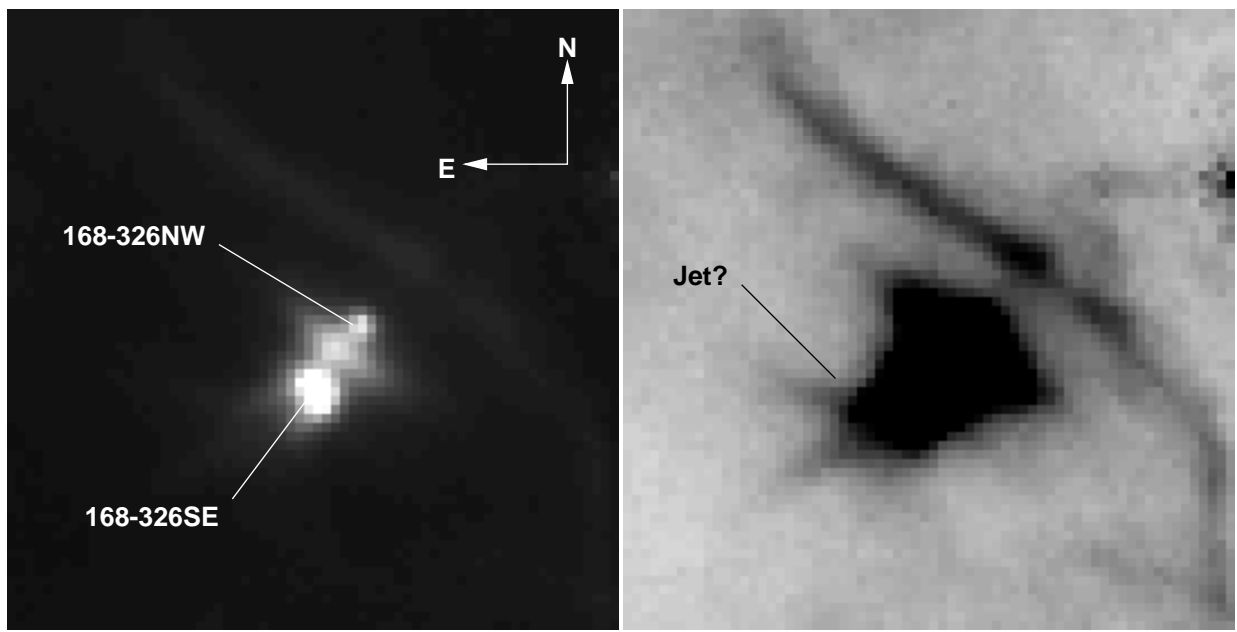}}
\caption{\oiii\ \emph{HST} image of LV 1 now resolved
into its binary proplyd components 168-326NW and SE. Left: Positive greyscale
showing the individual binary components. Right: Negative greyscale at
a different contrast to show fainter features, such as the proplyd
flow-stellar wind interaction bowshock and the predominantly
monopolar micro-jet with a PA of 85$^{\circ}$.}  
\end{figure}

\section{Radio/Optical Comparisons}

The identification chart for the Trapezium region in Figure 1 shows
$\theta^1$Ori A-F, the six LV objects, and the bowshock that marks
the interface between the low speed proplyd disk flow of LV 1 and 
the high speed stellar
wind of \thetc.  This interaction shields the LV 1 binary
system from a direct interaction with \thetc's wind and
happens on a much larger scale than the interaction between the two
proplyds which constitute LV 1 (168--326SE and 168--326NW).

Three distinct sources can be seen in the MERLIN 5 GHz radio contours
shown in Figures 4 and 5, which compare the 5 GHz radio emission to the
\ha\ and \oiii\ emission respectively.
%
\begin{figure}
\scalebox{0.9}
   {\resizebox{\textwidth}{!}
      {\includegraphics{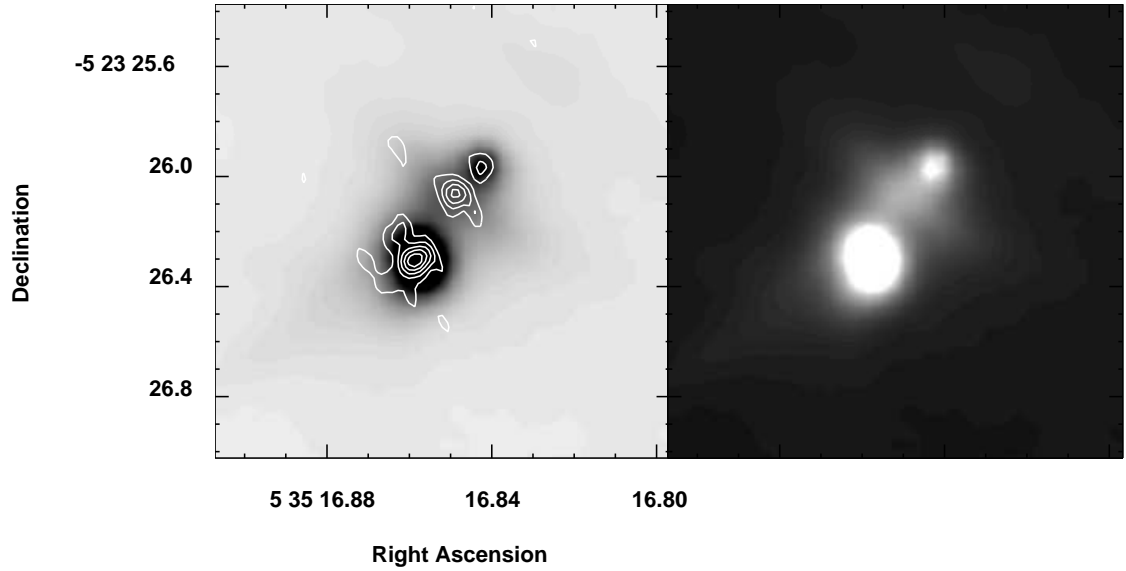}}}
\caption{Left: MERLIN 5 GHz contours at
$4\times10^{-4}\times(1,1.5,2,2.5,3)$ Jy/BEAM superimposed on the
negative greyscale \emph{HST} \ha\ (F656N) image of LV 1 (168--326SE and
326NW). Right: Greyscale \emph{HST} \ha\ (F656N) image of LV 1 (168--326SE
and 326NW). Note that this field-of-view is about the
size of the oval marked LV 1 in Figure 1.}
\end{figure}
%
%
\begin{figure}
\scalebox{0.9}
   {\resizebox{\textwidth}{!}
      {\includegraphics{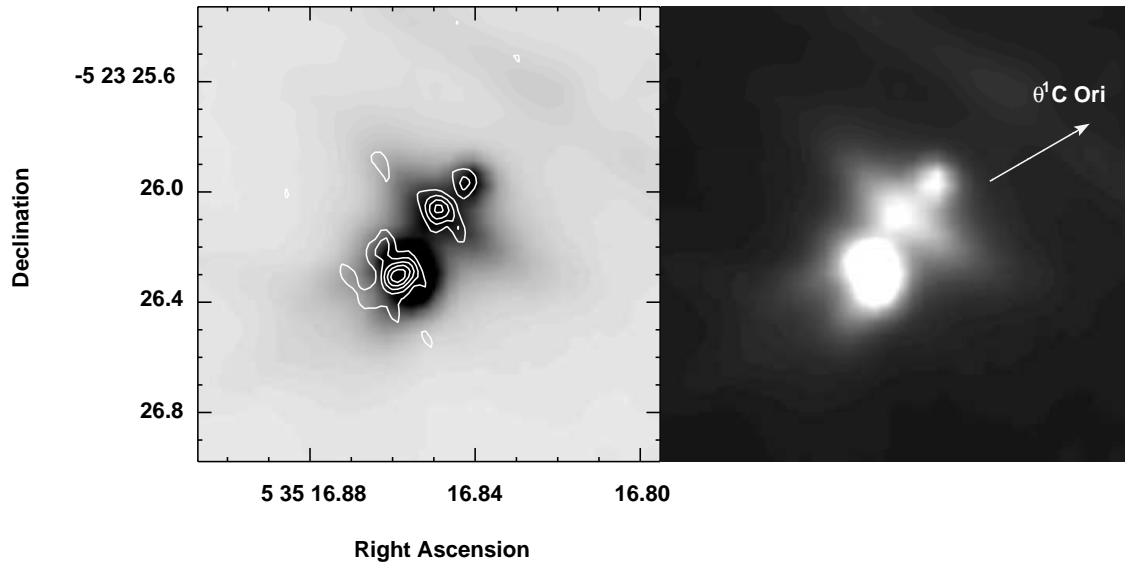}}}
\caption{As Figure 4 but versus the \oiii\ (F502N) \emph{HST} image,
along with the direction of the ionizing star \thetc.}
\end{figure}
%
The result of the alignment techniques described above is that the
south-east radio source corresponds to 168--326SE and the north-west
radio source corresponds to 168--326NW.  This leads to the strong
possibility that the radio source located between the binary
components is the result of an interaction between them.

This interaction zone is not as prominent in the \ha\ image.  
Furthermore, the peak radio to
peak \ha\ flux ratios for the two proplyds are the same, whereas this
ratio for the interaction source is just over 2.56 times that for the
proplyds. This could indicate that the interaction
radio emission has a non-thermal origin (see Section 5).

However, the interaction zone can be seen in the light of the \oiii\
line, as shown in Figure 4.  This could mean that the region has a
temperature high enough (i.e. $>10^{4}$ K)  to radiate predominantly
in [O{\sc iii}], as opposed to \ha. A different density in this
region compared with the proplyds, combined with this higher
temperature could therefore generate sufficiently enhanced thermal
radio emission (see $\S$ 5) to give the enhanced radio/\ha\ 
ratio that is observed without invoking non-thermal radio processes.
  
The position of the
brightest part of the interaction source in the \oiii\ image is of
particular interest.  In this image, the tail of 168--326SE can be seen
pointing almost exactly away from \thetc.  This is a common
feature of many of the proplyds (Bally et al. 1998) and leads to the
conclusion that the diffuse photon flux of the nebula is causing a
flow off the `back' of the proplyds. The brightest point of the
interaction source is on the opposite side of 168--326NW to \thetc\ and
not directly on the axis joining the two binary components. 
A simplistic  explanation for this is that the interaction is
happening between material flowing off the `back' of 168--326NW into
the flow from the `front' of 168--326SE.

\section{Modelling the Interaction}

The morphology of the binary proplyd system, LV 1, shown in
Figure 5 can be summarised as follows.  There is a bright interaction
peak of emission at 5 GHz between the two proplyds, which
is nearer the 
smaller proplyd, 168--326NW.  This peak is the brightest part of an arc
of emission in \ha\ and \oiii\ that is concave with respect to the
smaller proplyd.  The 
peak does not lie on the line connecting the two proplyd `heads', but
instead lies behind 168--326NW as seen from \thetc.  This
morphology seems to represent the first evidence for a hydrodynamic
interaction between two proplyds.  
A schematic model of such an
interaction is shown in Figure 6, in which the transonic
photoevaporated flows of a few 10's of\kms\ (as seen in Figure 2) from
each proplyd's disk, driven by the UV radiation from \thetc,
are colliding to form a dense interproplyd shell bounded by two weak
shocks.  
%
\begin{figure}
\centering
\scalebox{0.6}
   {\resizebox{\textwidth}{!}
      {\includegraphics{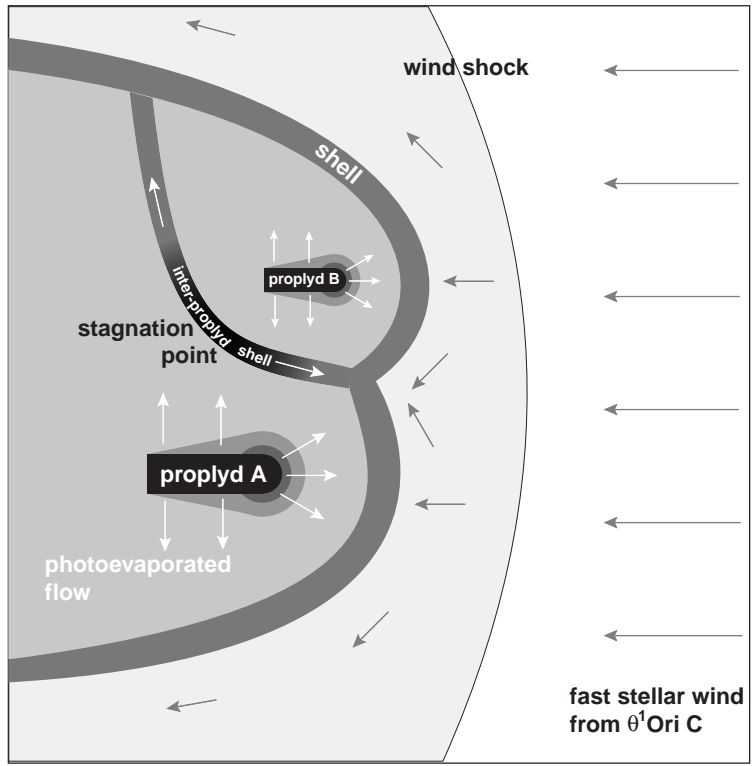}}}
\caption{A schematic of the three-flow interaction that will occur
when a binary proplyd system is found close to \thetc, showing the
plane containing the two proplyd major axes.  In addition to the
standoff shell between each proplyd and the stellar wind, there is
also an interproplyd shell.}
\end{figure}
%
The shell formed by the interaction of the disk flows from two proplyds
with the stellar wind of \thetc\ is also shown.
Canto, Raga, \& Wilkin (1996, hereafter CRW) have studied the
interaction between two  
hypersonic, non-accelerating, spherical winds forming an axisymmetric 
bowshock structure under the approximation that the interaction region 
forms a thin, momentum-conserving shell.  The situation with the
proplyds is more complicated, mainly because the individual flows are
non-isotropic, which will result in a totally asymmetric
bowshock.  However, the following principal results of the model
appear to have observational counterparts in the LV 1 system:
\begin{list}{\arabic{no1}.}{\usecounter{no1}\setlength{\rightmargin}{\leftmargin}}
\item There will exist a point, termed the \emph{stagnation point},
where the shell surface is perpendicular to the velocity vectors of
both flows.  This forms the `nose' of the bowshock and its position is 
determined by the condition of ram pressure balance; it will
be closer to the less powerful proplyd in terms of wind momentum
flux.  The normal to the shell surface at the stagnation point can be
considered the bowshock axis, even though it is not an axis of symmetry.
\item The shell will curve back towards the less powerful flow.  At
each azimuth around the bowshock axis, the shell will tend
asymptotically to a constant opening angle from the axis, but the
anisotropy of the flows will lead to this opening angle varying with
azimuth.
\end{list}
The stagnation point is the area of highest density in the interaction
shell and will therefore show up as a peak in the intensity of any
bowshock that lies between the proplyds and is therefore clearly
identifiable in Figure 5.  The anisotropy of the flows means that the
stagnation point need not necessarily lie on a line joining the proplyd
heads. The observed radio emission at the stagnation point
takes the approximate form of a point source, with some evidence of
extension at $PA\approx45^{\circ}$ in the lowest (3$\sigma$) contour
level (Figures 4 \& 5). If the radio emission does continue 
to extend in the same way
as the optical, it could be below the sensitivity of these
observations.

The details of deciphering the geometry of the interacting binary
system and then applying the model of CRW
is left to a subsequent paper (Henney 2001).  However, a
preliminary result of considering the 3-dimensional geometry of LV 1
leads to two possible geometrical configurations for the system, as
shown in Figure 7. 
%
\begin{figure}
\centering
\scalebox{0.6}
   {\resizebox{\textwidth}{!}
      {\includegraphics{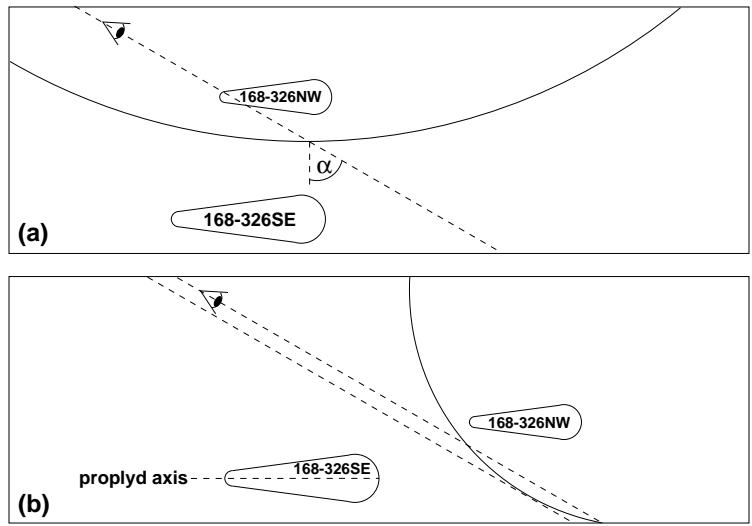}}}
\caption{Two possible geometrical configurations for the LV 1 binary
system.
\textbf{(a)}: Side-by-side configuration.
\textbf{(b)}: Head-tail configuration.  
The observer's line of sight
through the stagnation point of the shell is indicated by a dashed
line (in reality the line of sight is slightly inclined with respect
to the plane of the diagram). In configuration (a), the angle between
the line of sight and the bowshock axis, $\alpha$, is defined.
In configuration (b), the lower dashed line indicates the line of sight
that is tangent to the shell and the proplyd axis is defined.}
\end{figure}
%
Both configurations imply that the proplyds lie approximately between
the observer and \thetc, i.e. the observer is `behind' the
proplyds with respect to \thetc.  In configuration (a) the
proplyds are roughly side-by-side as seen from 
\thetc, with the stagnation point forming between them.
Whereas in configuration (b), the smaller proplyd is closer to
\thetc, resulting in the tail flow from this smaller proplyd
interacting with the head flow from the larger proplyd, with the
stagnation point forming close to the base of the smaller proplyd's
tail, as suggested in \S\ 3.  In both cases the stagnation point will
be seen to lie between the proplyds, but in
configuration (a) it will be superimposed on the tail of 168--326NW.
This superposition can be seen in the \emph{HST} image in Figure 5 and 
therefore favours the side-by-side configuration. 
However, this configuration places the observer within the `bowshock
cone', implying that an arc would not be observed.  The observed arc
in the \emph{HST} images favours configuration (b).  If the
observer does lie outside the bowshock cone, then the shell would be
seen tangentially at some point and the fact that it is clearly
resolved with the \emph{HST} would put a lower limit on its thickness.

In order to conclusively choose between these two configurations, the
morphological appearance of the shell has  been modelled for both
cases.  For this, a model for the shape of the shell was needed,
and has been provided by the CRW model.  The shape of a CRW shell is
characterised by its asymptotic opening angle, $\theta_{\infty}$, which
is uniquely determined by the momentum flux ratio, $\beta$, of the two 
interacting flows.  In Figure 8, this shape for
$\beta=5$, $\theta_{\infty}=61.9^{\circ}$ has been used to calculate
the appearance of a shell at various viewing angles, $\alpha$, the
angle between the line of sight and the bowshock axis for a
cylindrically symmetric bowshock.
%
\begin{figure}
\resizebox{\textwidth}{!}
   {\includegraphics{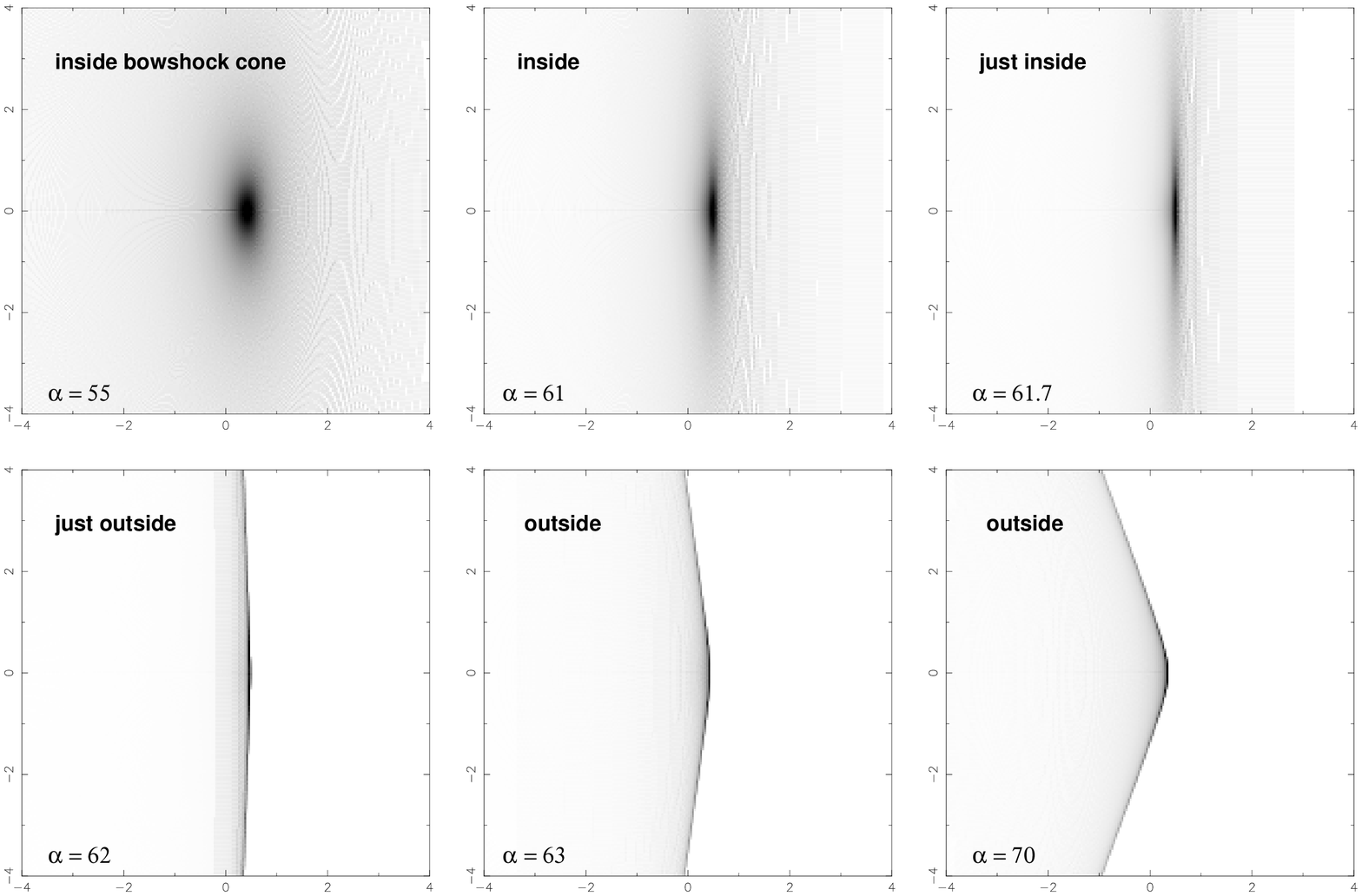}}
\caption{Modelled images of a cylindrically symmetric CRW shell for an
asymptotic opening angle, $\theta_{\infty}=61.9^{\circ}$ and a 
wind momentum flux ratio, $\beta=5$ at various viewing angles, $\alpha$.
Whether the observer is inside or outside the bowshock cone is
indicated. The greyscale intensity in each panel is normalized to the
maximum pixel of the image.}
\end{figure}
%
It can be seen that the morphology of the modelled shell when viewed
from within the bowshock cone with $\alpha=61$\degree is close to that
of the observed shell, but this modelled shell has no apparent
curvature.  In fact, the observed curvature is similar to a modelled
shell with $\alpha=63^{\circ}$, with the observer outside the bowshock
cone.  This discrepancy could be explained by the fact that the
bowshock need not be cylindrically symmetric. Each proplyd has a
distinctive cometary shape extending away from \thetc\ along an axis
referred to as the proplyd axis (see Figure 7). If there is a
contribution to a proplyd's
flow from the entire length of the proplyd axis, the bowshock between
two proplyds is likely to be flattened in the direction
perpendicular to the proplyd axis.  The appearance of an  
asymmetric bowshock can be modelled by arbitrarily decreasing $\beta$
and therefore increasing $\theta_{\infty}$ in the
direction parallel to the proplyd axis.  This will result in a
bowshock whose cross-section resembles an ellipse with a major axis
parallel to the proplyd axis. Figure 9 is a set of modelled images for
such an asymmetric bowshock, with the $\theta_{\infty}$ increased to
75\degree in the direction parallel to the proplyd axis.
%
\begin{figure}
\resizebox{\textwidth}{!}
   {\includegraphics{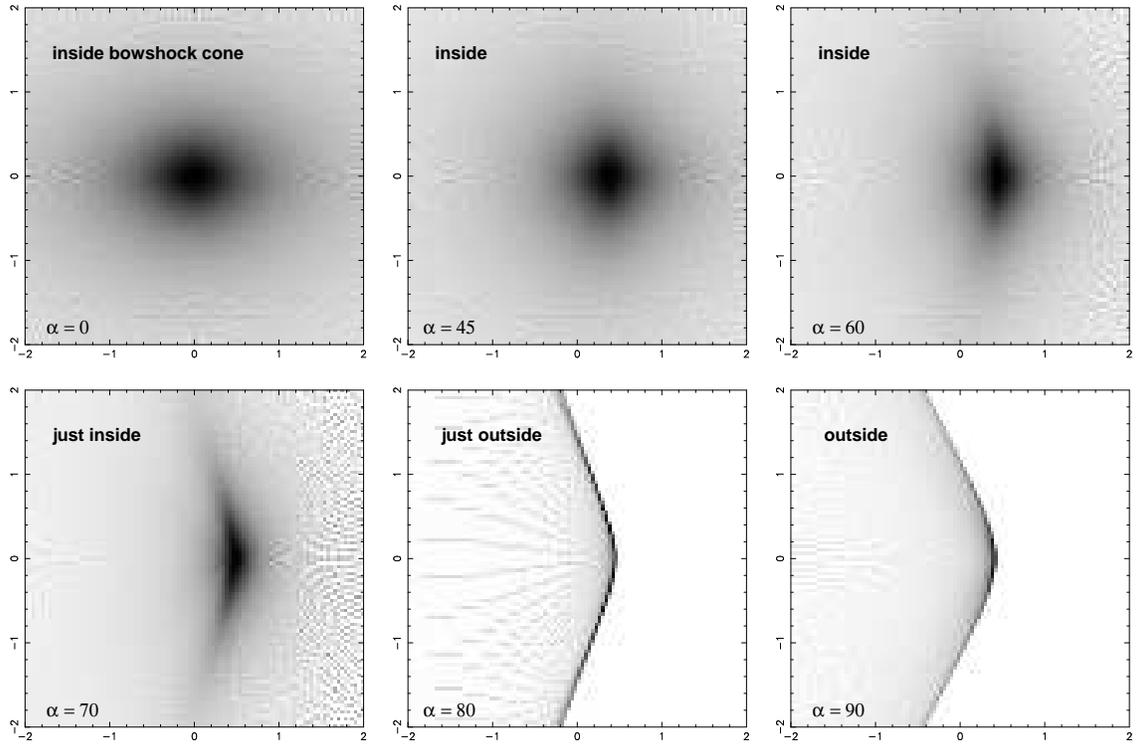}}
\caption{Modelled images of an asymmetric CRW shell for orthogonal
asymptotic opening angles, $\theta_{\infty}=75^{\circ},~62^{\circ}$
and orthogonal wind momentum flux ratios, $\beta=2,~5$ at various
viewing angles, $\alpha$.  Whether the observer is inside or outside
the bowshock cone is indicated. The greyscale intensity in each panel
is normalized to the maximum pixel of the image.}  
\end{figure}
%
With $\alpha=0$ the intrinsic flatness of the bowshock can be seen,
giving rise to elliptical isophotes with a major axis in the
horizontal direction.  As $\alpha$ increases this horizontal
elongation decreases and at $\alpha\approx45^{\circ}$, the intrinsic
elongation has been cancelled out, giving rise to approximately
circular isophotes.  As $\alpha$ continues to increase the apparent
elongation is in the vertical direction and some apparent curvature is
visible along the bowshock, even with the observer still inside the
bowshock cone.  With $\alpha=70^{\circ}$, the morphology of the
modelled image is a good match for the observed morphology, in terms
of both  smoothness and curvature (see Figures 4 \& 5).  
In these modelled images (Figures 8 \& 9), the
thickness of the shell has been assumed to be unresolved, as can be
seen in those images where the shell is viewed tangentially.  Henney
(2001) predicts that the shell's thickness is likely to be unresolved
with the \emph{HST}.  However, the presence of a magnetic field in the 
shell will act as an additional pressure, which will thicken the
interproplyd shell.  If this magnetic pressure dominates the
thermal pressure in holding the shell up against the ram
pressure of the two incident flows, the density in the region will be
lower than that predicted by Henney (2001).  A lower density will
lower the cooling rate and cause the cooling length to increase.
Therefore a significant magnetic field increases both the overall size 
of the shell and the cooling length within it.  If this effect is
large enough, the shell could be viewed from outside the bowshock cone,
and therefore tangentially, and still be resolved. 
Dust (other than foreground
extinction from the neutral veil) will be unimportant because the
column density of the shocked shell is too low.

\section{Nature of the Radio Emission}

An interaction zone between the two components of the binary proplyd,
LV 1, has been detected in both the radio and optical domains.  This
appears as an extended arc, containing a compact (5 GHz) knot. The
morphology of the interacting proplyd binary is 
similar to that of other interacting binaries such as WR 147 (see
e.g. Williams et al. 1997; Niemela et al. 1998).  In such cases the
interaction radio source is non-thermal, synchrotron emission caused by
shock-accelerated, relativistic electrons.  High energy electrons can
be produced in this scenario, because at least one of the winds has a
speed of order $1000 \kms$.  In the case of the proplyds, the speed
of the evaporated flows is thought to be a
few times the sound speed of $10\kms$ and the bulk of the emission from
LV 1 appears to be from gas flowing at velocities of less
than $50\kms$ (Figure 2).  Two interacting flows each with this
velocity, would lead to two shocked layers at a post-shock
temperature of $\approx3\times10^{4}$ K (Dyson \& Williams 1997).
Although this is high enough to explain the enhanced [O{\sc iii}]
emission from the interaction region, it is not sufficient to account
for the existence of relativistic electrons.  Therefore, synchrotron
emission is not expected from these low speed collisions.

The 5 GHz radio/\ha\ intensity ratio is approximately 2.6 times higher
for the interaction region than for the proplyds.  For identical
physical conditions in the proplyds and the interaction region, this
could be  indicative of non-thermal radio emission from the
interaction zone. However, it has already been established that the
post-shock material in this zone is likely to 
have a higher temperature than the proplyds.  Post-shock material is
also unlikely to have the same density as the unshocked material
flowing from the proplyd.  The canonical density jump for a strong
shock is a factor of 4 (Dyson \& Williams 1997), but for weaker shocks 
this factor is lower.  Once the post-shock material has cooled, the final
compression ratio as compared with preshocked material will be higher
than this. It should also be
noted that the material is flowing from the proplyd down a density,
$N$, gradient that, 
for an accelerating spherical flow, would scale with radial 
distance from the centre of
the proplyd disk, $r$, as follows, 
\begin{equation}
N(r)=N(r_{{\rm if}})\left(\frac{r_{{\rm if}}}{r}\right)^{2}
\left(\frac{V_{{\rm if}}}{V(r)}\right)
\end{equation}
where $r_{{\rm if}}$ is the radius of the proplyd ionization front
(IF), $N(r_{{\rm if}})$ is the density at this radius,  $V_{\rm if}$
is the gas velocity at the IF ($\approx$ sound speed) and 
V({\rm r}) is the gas velocity.  The interaction
between 168--326SE and 168--326NW is happening at approximately 2-3
IF radii.  Therefore, in the case of a spherical flow
from the proplyds, a density drop of a factor of $>4$ would be expected 
in the pre-shocked proplyd flow as compared with material flowing from 
the IF.  As a result, even after compression due to the
shock, the density in the post-shock material in the interproplyd
shell is likely to be lower than the density at the proplyd's
IF.  This difference in both temperature and density
between the IF and the interproplyd shell could provide
an explanation for the observed difference in 5 GHz radio/\ha\
intensity ratio.  However, the contribution from the high temperature
gas, which will rapidly cool through [O{\sc iii}] emission, depends on
its cooling length as compared with the total length of the
interproplyd shell.  If this contribution from the post-shock cooling
layer is insignificant, the majority of the material in the shell will 
be at the equilibrium nebular temperature of $10^4$ K.  The density of 
this cooled material will be higher than that in the cooling layer,
but still different to that at the proplyd IF.

Using the tabulated values of $j_{\rm H\alpha}$, the emission
coefficient of the \ha\ line provided by Storey \& Hummer (1995) at
various temperatures, $T$, the \ha\ flux density from an object with
uniform density, $N$, at a distance, $D$, for case B recombination is
\begin{equation}
F({\rm H\alpha})=1.850\times10^{-21}T^{-0.932}N^{2}R^{3}D^{-2}~{\rm(erg~cm^{-2}~s^{-1}),}
\end{equation}
where $R$ is the radius of the object. Using typical values for the
proplyds, equation (\theequation) predicts 
flux densities of order $10^{-11}~{\rm erg~cm^{-2}~s^{-1}}$, which is
consistent with the total fluxes calculated by O'Dell (1998).

The radio flux, $S_{\nu}$, due to thermal brehmstrahlung from a source
observed at a frequency, $\nu$, is
\begin{equation}
S_{\nu}=\frac{2k\nu^{2}\Omega}{c^{2}}T\left(1-e^{-\tau_{\nu}}\right),
\end{equation}
where $\Omega$ is the solid angle of the source and $\tau_{\nu}$ is the
optical depth.  This is reasonably approximated by the following
relationship 
\begin{equation}
\tau_{\nu}=8.235\times10^{-2}\left(\frac{T}{\rm
K}\right)^{-1.35}\left(\frac{\nu}{\rm GHz}\right)^{-2.1}\left(\frac{EM}{\rm pc~cm^{-6}}\right)
\end{equation}
where $EM\approx N^{2}R$ is the emission measure (Mezger \& Henderson
1967).  For typical values for the proplyds at 5 GHz, this is close to
1, so the proplyds can neither be treated as optically thin nor thick
at this frequency.  A full treatment allowing for an arbitrary optical
depth is required. 

Therefore, the peak \ha\ flux ratio for two objects at
temperatures, $T_{1}$ and $T_{2}$, densities, $N_{1}$ and $N_{2}$ and
sizes $R_{1}$ and $R_{2}$ is given by
\begin{equation}
\frac{F({\rm H\alpha})_{1}}{F({\rm
H\alpha})_{2}}=\left(\frac{T_{1}}{T_{2}}\right)^{-0.932}\left(\frac{N_{1}}{N_{2}}\right)^{2}\left(\frac{R_{1}}{R_{2}}\right)
\label{ha_ratio}
\end{equation}
and the 5 GHz radio flux ratio for the two objects with optical
depths, $\tau_{\rm 5GHz(1)}$ and $\tau_{\rm 5GHz(2)}$ is given by
\begin{equation}
\frac{S_{\rm 5GHz(1)}}{S_{\rm
5GHz(2)}}=\frac{T_{1}\left(1-e^{-\tau_{\rm5GHz(1)}}\right)}{T_{2}\left(1-e^{-\tau_{\rm 5GHz(2)}}\right)}.
\label{5ghz_ratio}
\end{equation}
These ratios are known for 168--326SE compared to the interaction zone 
from the \emph{HST} and MERLIN \mbox{5 GHz} data, to be 3.4 in \ha\ and
1.3 in the radio.  Equations (5) and (6) can be solved
simultaneously for $T_{2}$ and $N_{2}$, given values for $T_{1}$ and
$N_{1}$, i.e. given a temperature and density for 168--326SE a
temperature and density consistent with the flux ratios can be found
for the interaction zone.  A scale-size is needed in order to
estimate the volume of material being observed.  For both
the proplyd and the interaction zone, this
was chosen to be $10^{-4}$ pc ($\equiv20$ AU), which is approximately
the radius of the IF for a proplyd.  With $T_{1}=10^{4}$
K and $N_{1}=10^{6}~{\rm cm^{-3}}$ for the proplyd, the equations are
solved with $T_{2}=25470$ K and $N_{2}=0.839\times10^{6}~{\rm
cm^{-3}}$ for the interaction zone.  These values are consistent with
both the temperature predicted by the shock speed and the expectation
that the density should be lower in the interaction zone.
However, the values are sensitive to the chosen scale-size, initial
temperature ($T_{1}$) and density ($N_{1}$). The \ha\ fluxes could also 
be significantly affected by extinction.  Nevertheless,
this treatment shows that by changing the temperature and density in
the post-shock region, the required flux ratios can be arrived at
without the need for a non-thermal source of radio emission, as long
as there is sufficient material at the high post-shock temperature and
lower density. If there is insufficient post-shock material, as
suggested by  Henney  (2001), the difference in density between the
proplyd and the cooled material could still explain the difference in
5 GHz radio/\ha\ intensity ratio between the two regions. The
intensity ratio is also affected by the physical size of the emitting
region, so the uncertainty in the shell thickness and cooling length
makes it difficult to determine whether the radio emission has a
thermal origin.

The spectral index of the interaction compact radio source is now
needed to distinguish conclusively between a thermal or non-thermal
origin.  For this, observations are needed at 1.5 GHz (20 cm) with a
similar angular resolution  to that of the 5 GHz MERLIN observations
presented here.

\section{Conclusions}

The Orion proplyd originally designated LV 1 is shown at both
radio (with MERLIN) and optical wavelengths (with the HST) 
to be a proplyd binary system
with an interaction zone separating the two components
168-326NW and 168-326SE.

Comparison of detailed hydrodynamical modelling of the interaction
zone with the observations has identified the geometry of the system 
convincingly.

A strong 6 cm radio source coincides with the stagnation
point in this interaction zone. The present observations cannot
distinguish between a thermal or non-thermal origin
for this radio emission. Observations at 20 cm but with 
a similar angular resolution (0.1 \arcsec) are awaited to clarify this point
through measurement of the radio spectral index.
  
The existence of a  high-speed micro-jet from the binary component
168-326SE 
is also suggested by spectral observations
and HST imagery. 

\vspace{10mm}

\noindent {\bf Acknowledgements}

MFG would like to acknowledge a studentship received from the
Particle Physics and Astronomy Research Council (PPARC).  MERLIN is a
national facility operated by the University of Manchester on behalf
of PPARC in the U.K. The work presented in this paper is based on
observations made with MERLIN and 
the NASA/ESA Hubble Space Telescope, the latter obtained
from the data archive at the Space Telescope Science Institute. STScI
is operated by the Association of Universities for Research in
Astronomy, Inc. under NASA contract NAS 5-26555.

\end{document}